\documentclass[twocolumn,showpacs,preprintnumbers,amsmath,amssymb]{revtex4}
\usepackage{graphicx}
\usepackage{dcolumn} 
\usepackage{bm}      
\usepackage[mathscr,mathcal]{euscript}
\usepackage{eufrak}
\begin{document}

\title{Global phase diagram of three-dimensional extended Boson Hubbard
  model -- a continuous time Quantum Monte Carlo study}

\author{Bin Xi$^{1,4}$, Fei Ye$^{2}$, Weiqiang Chen$^{3}$, Fuchun
  Zhang$^{3}$ and Gang Su$^{1}$ }

\email[]{Email: gsu@gucas.ac.cn}

\affiliation{$^{1}$College of Physical Sciences, Graduate University
of  Chinese Academy of Sciences, Beijing 100049, China}

\affiliation{$^{2}$College of Material Sciences and Optoelectric Technology,
  Graduate University of Chinese Academy of Sciences, Beijing 100049,
  China}

\affiliation{$^{3}$Centre of Theoretical and Computational Physics
and Department of Physics, the University of Hong Kong, Hong Kong,
China}

\affiliation{$^{4}$Institute of Theoretical Physics,
Chinese Academy of Sciences, Beijing 100190, China}

\date{\today}

\begin{abstract}
  We present the global phase diagram of the extended boson Hubbard
  model on a simple cubic lattice by quantum Monte Carlo simulation with
  worm update algorithm. Four kinds of phases are supported by this
  model, including superfluid, supersolid, Mott, and charge density wave
  (CDW) states, which are identified in the phase diagram of chemical
  potential $\mu$ versus nearest neighbor interaction $V$.  By changing
  the chemical potential, a continuous transition is found from the Mott
  phase to a superfluid phase without breaking the translational
  symmetry. For an insulating CDW state, adding particles to it gives
  rise to a continuous transition to a supersolid phase, while removing
  particles usually leads to a first-order one to either supersolid or
  superfluid phase. By tuning the nearest neighbor interaction, one can
  realize the transition between two insulating phases, Mott and CDW
  with the same particle density, which turns out to be of the
  first-order.  We also demonstrate that a supersolid phase with average
  particle density less than 1/2 can exist in a small region of $\mu-V$
  phase diagram.
\end{abstract}
\maketitle

\section{Introduction}
Lattice models of interacting bosons and fermions such as Hubbard model
and its various generalizations are usually strongly correlated systems
exhibiting various phases with competing orders, which are of
fundamental interest in fields of both condensed matter and cold atomic
physics. Interests on both types of Hubbard models are renewed recently,
since they can be realized in cold atomic gases loaded in optical
lattices (for a review see Refs.~[\onlinecite{bloch2008}] and
[\onlinecite{zhai2009}] and references therein). Unlike fermions, there
is a natural superfluid order for free bosons at zero temperature driven
by the kinetic energy. When the interaction is switched on, the bosons
are likely to be localized in various crystalline patterns, which may
coexist with superfluid order
\cite{batrouni1995,otterlo1995,sengupta2005,batrouni2006} to give a
realization of intriguing ``supersolid'' state that has been pursued for
decades since 1950s
\cite{penrose1956,andreev1960,chester1970,leggett1970}. Recently, people
have observed the non-classical rotational inertia in solidified $^4$He
\cite{kim2004a,kim2004b} implying a possible supersolid state, which, in
spite of the controversy over this topic, also triggers extensive
studies on various boson Hubbard models.

Experimentally, the boson Hubbard model can be used to mimic the
granular superconductors, where the Cooper pairs are described as
bosons, which has been studied by Fisher \textit{et al.}
[\onlinecite{fisher1989}] two decades ago, where with only on-site
repulsive interaction they showed that bosons can form either Mott
insulating state with integer filling or superfluid state. Recent
experimental progress in cold atomic system provides another
realization of boson Hubbard model by loading atoms into an optical
lattice with possible long range interactions through dipole
interaction \cite{griesmaier2005,rieger2005,ni2008}, or mediated by
other intermediate states or fermions
\cite{buchler2003,scarola2005,henkel2010}. In addition, the boson
models also share similarities with quantum magnets, e.g., the
uniaxial magnetization corresponds to insulating states of boson
Hubbard model (e.g. Ref. \onlinecite{liwei2010}), while the
easy-plane magnetization corresponds to the superfluid state. Hence,
the studies on the boson Hubbard model may shed light on some common
issues of strongly correlated lattice models.

Generally speaking, boson models with interactions at zero temperature
have two principal phases: (i) the superfluid and (ii) the
incompressible insulating state, which are favored respectively by
kinetic and interaction energies, and can coexist. Depending on the
features of interaction terms, there are several types of insulating
phases, such as Mott, valence bond crystal, and charge density wave
(CDW). Note that we in this article define the incompressible states
with oscillating density profile as CDW, though the bosons may not carry
charges.

The extended boson Hubbard (EBH) model with onsite ($U$) and nearest
neighbor ($V$) interactions is a minimal model in favor of CDW and
supersolid phases, which has the form of
\begin{eqnarray}\label{hmt}
  \hat{H}&=&-t\sum_{\langle
    i,j\rangle}(\hat{b}^{\dagger}_i\hat{b}_j+\hat{b}^{\dagger}_j
  \hat{b}_i)+\frac{U}{2}\sum_i\hat{n}_i(\hat{n}_i-1)\nonumber\\
  &&+V\sum_{\langle i,j\rangle}\hat{n}_i\hat{n}_j-\mu\sum_i\hat{n}_i,
\end{eqnarray}
where $\hat{b}^{\dagger}_i$ ($\hat{b}_i$) is the creation (annihilation)
bosonic operator at site $i$, $t$ is the hopping amplitude,
$\hat{n}_i=\hat{b}^{\dagger}_i\hat{b}_i$ is the particle number, $\mu$
is the chemical potential, and $\langle i,j\rangle$ runs over all
nearest neighbors. Recently, Hamiltonian Eq.~\eqref{hmt} and its
hard-core version (equivalent to the quantum spin-1/2 XXZ model) with
different underlying lattices have been extensively studied in different
parameter regimes
\cite{kuklov2004,sengupta2005,melko2005,batrouni2006,lu2006,
  yi2007,iskin2009,yamamoto2009,pollet2010,ueda2010}. However, a global
phase diagram of the three-dimensional (3D) EBH model [Eq.  \eqref{hmt}]
is still of lack. As there is no sign problem for the EBH model, the
quantum Monte Carlo (QMC) simulation is the most convenient tool for
this purpose. The worm algorithm
\cite{prokof1996,prokof1998,kashurnikov1999} will be invoked to study
Hamiltonian \eqref{hmt} on a simple cubic lattice, together with other
perturbation and mean-field approaches.

The system described by the EBH model can give rise to a charge ordered
crystal at commensurate fillings. The first one is for half filling
$\rho=0.5$, and the corresponding solid state is labeled as CDW
I. Doping \emph{particles} into this state can lead to a supersolid
state \cite{sengupta2005,batrouni2006}. However, as shown in
Ref. \cite{sengupta2005}, doping \emph{holes} into it acts quite
differently, which may not result in a supersolid state with $\rho<0.5$,
but a phase separation between superfluid and CDW I states, which signals
a first-order phase transition. Their argument is based upon the
following two observations. (I) Taking one particle out of a perfect CDW
crystal with half filling costs almost no potential energy, but only
chemical potential. At the same time, the hopping hole also gains a
kinetic energy which is quadratic in $t$ $(\sim t^2)$. For a perfect CDW
crystal, these three processes are balanced, so one cannot take one
particle out. (II) The CDW phase breaks the translational symmetry,
leading to a two-fold degenerate ground state. If holes are doped into
the domain wall between these two degenerate phases, the kinetic energy
gained is proportional to $t$. Hence, the CDW phase is unstable toward
the domain wall formation if the hole density exceeds $L^{-1}$ for $L^d$
lattice, though it is still stable against losing one particle. This
argument perfectly explains the first-order phase transition from the
CDW I to superfluid state with $\rho\le 0.5$, but it fails in two
circumstances. The first is that in one dimension the kinetic energy is
always linear in $t$, and the corresponding transition is of the
Kosterlitz-Thouless type \cite{batrouni2006}. The other is that if $V$
is comparable to $t$ the kinetic energy of holes is also linear in $t$,
which may result in the supersolid phase with the particle density less
than half filling (see Sec.~\ref{sec:case-finite-t}). This can be
verified by the mean-field calculations \cite{yamamoto2009,ye2010}.

At unitary filling, the ground state can be either a uniform Mott
insulator with one particle per site or a charge ordered crystal with
two particles on one sublattice and empty on the other one which is
labeled as CDW II. There is a critical region around $U\sim zV$, where
the two states with different translation symmetries become degenerate,
and however, they are separated thermodynamically, i.e., any local
perturbation cannot take one to the other. Correspondingly, the
transition between them is a first-order one. It is noted that the
aforementioned transition from the superfluid to CDW I state by tuning
the chemical potential is of the \emph{weak} first-order
\cite{kuklov2004}. Far less attention has been paid to the region with
$zV\sim U$ by now, of which the details are given as part of the phase
diagram in this article. To plot the ground state phase diagram, we
focus on the case with small hopping and average particle density around
or smaller than 1. For larger $t$ or $\rho$, we expect no essentially
new physics.

This article is organized as follows. In
Sec.~\ref{sec:ground-state-phase}, we shall present the global phase
diagram. The details of the order parameters will be discussed in
Sec.~\ref{sec:order-parameters}. The conclusion will be given in last
section.

\section{Global Phase Diagram}
\label{sec:ground-state-phase}
\subsection{Classical case with $t=0$}
We start from the classical case without hopping. The energy per site of
ground state is a function of the particle numbers on the two
sublattices, $n_A$ and $n_B$,
\begin{eqnarray}
\label{eq:1}
\epsilon^{(0)}(n_A,n_B)&=& -\frac{\mu}{2}(n_A+n_B)  +
\frac{zV}{2}n_An_B \nonumber\\
&&+ \frac{U}{4}[n_A(n_A-1) +n_B(n_B-1)]\;,
\end{eqnarray}
where the coordination number $z=6$ for the simple cubic lattice. The
states can be labeled by $(n_A,n_B)$. The Mott states correspond to
$n_A=n_B$, and the CDW states with $n_A\ne n_B$ break the translational
symmetry, which is two-fold degenerate. In this article, we define the
state $(1,0)$ as CDW I state, and $(2,0)$ as CDW II state and we only
consider $n_A>n_B$ for the CDW states for convenience.
\begin{figure}[htbp]
\centerline{\includegraphics[width=6.0cm]{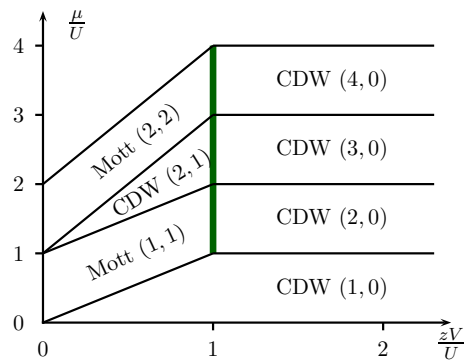}}
\caption[]{\label{fig:zerot} (Color online) The phase diagram with zero
  hopping $t=0$, where the states are labeled by the particle numbers on
  two sublattices $(n_A,n_B)$ with the assumption $n_A\ge n_B$. The
  states with $n_A<n_B$ can be obtained by inversion. The CDW states
  with one sublattice empty on the right side. Some states can only
  exist on the green thick solid line, e.g., the states $(3,1)$ and
  $(1,3)$.}
\end{figure}

For $\mu<0$ the ground state is a vacuum without any particles. As the
chemical potential is increased $(\mu>0)$, the particles are loaded into
one sublattice to form a charge ordered pattern with $n_A=1$ and
$n_B=0$, i.e., CDW I state. If we further increase the chemical
potential, more particles are loaded into the cubic lattice, which fill
either the empty sites if $zV/U<1$ to form a uniform Mott state, or the
occupied sites if $zV/U>1$ leading to a CDW II state. In the Mott state,
each particle interacts with its nearest neighbors, which effectively
lowers the chemical potential to be $\mu-zV$, and then the critical line
between CDW I and Mott states is simply $\mu=zV$. While that between CDW
I and II states is a horizontal line $\mu=U$ because the chemical
potential only needs to compensate the on-site interaction $U$ for
adding new particles.

Similarly, by studying the instability of adding particle to a state
$(n_A,n_B)$, one can determine all the phase boundaries between
different classical insulating states, as shown in Fig.
\ref{fig:zerot}.  There is a special vertical line $zV/U=1$, on which
many states can coexist with the same free energy. For example, on the
boundary between Mott and CDW II states, there are actually three
macroscopic states, which are (1,1), (2,0) and (0,2).  In fact, some of
them only exist on this line in the absence of hopping terms,
e.g. states (3,1) and (1,3) on the boundary between Mott (2,2) and CDW
(4,0) states. These degenerate states are macroscopic which cannot be
transformed to each other smoothly by local perturbations, i.e., there
are infinitely high barriers between these macroscopic states.

\subsection{Case for finite $t$}
\label{sec:case-finite-t} 
\begin{figure}[htbp]
\centerline{\includegraphics[width=8cm]{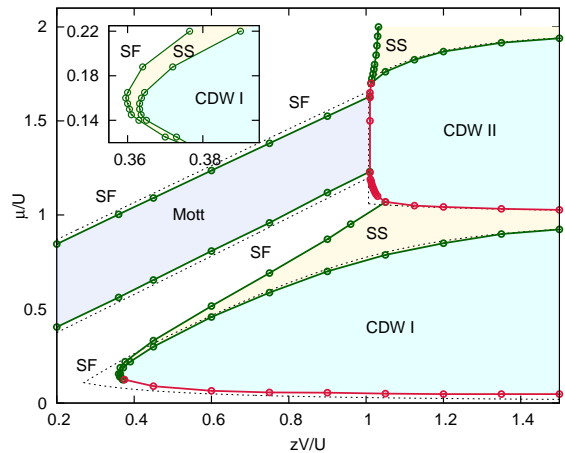}}
\caption[]{\label{fig:phasediagram}(Color online) The phase diagram in
  the plane of $\mu/U$ \emph{vs}. $zV/U$ for the extended Bose Hubbard
  model on a simple cubic lattice, where $U=40t$ and the lattice size is
  $12\times 12\times 12$. Four kinds of phases including the superfluid
  (SF), supersolid (SS), Mott and CDW states are identified.  The CDW
  states are further classified as I and II by the filling numbers. The
  solid lines with circles are the phase boundaries calculated by QMC
  simulations, and the dotted lines are from the perturbation
  calculations. The green lines are the second-order phase boundaries,
  and the red ones are of the first-order. The inset is a zoom near the
  tip of CDW I lobe, where one can find a narrow region below the lobe
  corresponding to a supersolid phase with the filling less than one
  half. For details of order parameters one may refer to
  Figs.~\ref{fig:extrapolation}, ~\ref{fig:fixedmu} and
  ~\ref{fig:fixmurho}.}
\end{figure}

For the case with a finite hopping $t$, the
particles (holes) adding to an insulating state can gain a kinetic
energy, which results in the shrinking of insulating areas in the phase
diagram (Fig.~\ref{fig:phasediagram}) comparing with the classical case
(Fig.~\ref{fig:zerot}). In three dimensions, these mobile bosons
condense at low temperature leading to the superfluidity, which enriches
the phase diagram. There are four phases for the EBH model characterized
by the following three quantities: the particle density $\rho$,
superfluid density $\rho_s$, and static structure factor $S_{\vec{\pi}}$
at momentum $\vec{\pi}=(\pi,\pi,\pi)$ with
\begin{eqnarray}
\rho_s &=& \langle W^2_x+W^2_y+W^2_z\rangle/(2\beta m),  \nonumber\\
S_{\vec{\pi}}&=& \frac{1}{N^2} \sum_{\vec{r},\vec{r}'}e^{-i\vec{\pi}\cdot
(\vec{r}-\vec{r}')}\langle n_{\vec{r}} n_{\vec{r}'} \rangle,
\end{eqnarray}
where $W_{x,y,z}$ are the winding numbers along $x,y$ and $z$
directions, $\beta$ is the inverse temperature, and $m=2/t$ is the
effective mass of the bosons. In the insulating Mott and CDW states,
particles are localized by the interaction and the local particle number
is quantized as in the classical case. The pure superfluid state has
nonzero superfluid density $\rho_s$ and a vanishing static structure
factor $S_{\vec{\pi}}$, while both are finite in the supersolid phase.

Fig.~\ref{fig:phasediagram} is the phase diagram determined by the QMC
simulation with the worm update algorithm, where the solid lines with
circles are the QMC results and the dotted lines are from the
perturbation expansion in the strong coupling limit \cite{iskin2009}
where the insulating states become unstable against adding or removing
particles. It is seen that the perturbation results agree quite well
with those of the QMC simulation in part of the phase diagram, but it is
still not applicable in some regions since it cannot deal with the
superfluid order.

Comparing with the classical case in Fig.~\ref{fig:zerot}, the CDW I
state is detached from its insulating neighbors, i.e., the Mott, CDW II
and vacuum states. The upper boundary of vacuum state is actually
lowered below $\mu=0$ due to the hopping of bosons, which is not shown
in Fig.~\ref{fig:zerot}. The gaps between different insulating states
are filled with the superfluid and supersolid phases. The lower boundary
of CDW I state is a critical line on which there occurs a phase
separation between the superfluid and CDW I phases, which breaks the
U(1) gauge symmetry and the translational symmetry, respectively. The
transition between them is of the weak first-order across which the
particle density and superfluid density have a jump. Considering the
correspondence between the EBH model and spin models, this transition is
similar to the spin-flopping process in the two- or three-dimensional
anisotropic XXZ model in the presence of magnetic field pointing in
$z$-axis, which is equivalent to the EBH model in the hard-core limit
\cite{masanori1997,batrouni2000,schmid2002}. As explained in Ref.
[\onlinecite{sengupta2005}], this first-order phase transition with a
particle number jump is due to the fact that the CDW I phase is unstable
toward the domain wall formation if the filling number exceeds $L^{-1}$
though it is still stable against doping one hole.

Doping particles upon the CDW I phase by increasing the chemical
potential does not lead to a first-order transition as in the hard-core
EBH model (or equivalently the XXZ model), where the particle-hole
symmetry makes the upper and lower boundaries of CDW I phase
identical. In case of the soft-core bosons, these additional particles
can move upon the alternating charge ordered background with the
effective hopping amplitude $t^2$, which can Bose condense at zero
temperature without destroying the staggered density order, and thus
leads to a supersolid state with $\rho>0.5$. This transition is of a
second-order, as shown by a green solid line in
Fig.~\ref{fig:phasediagram} where all the second-order phase boundaries
are in green that are distinct from the first-order ones which are
colored in red.

Continuously increasing the chemical potential upon supersolid phase,
two different situations occur. (1) For $zV<U$, the particles like to
occupy the empty sites which weakens the CDW order accompanied by the
occurrence of superfluid order. Until some critical filling $\rho<1$,
the CDW order is completely destroyed and a pure superfluid state
appears with the translational symmetry restored. The transition is of
the second-order. (2) For $zV>U$, the additional particles are added to
the occupied sites so that the CDW order is actually enhanced until
finally entering into another insulating state, CDW II, through a
\emph{first}-order phase transition, of which the reason is the same as
that from the superfluid with $\rho<0.5$ to CDW I. In this sense, the
staggered density order in supersolid phase is inherited from the CDW I
state, not related to the CDW II state, since doping holes to the CDW II
state cannot \emph{smoothly} result in the supersolid state.

\begin{figure}[htbp]
\centerline{\includegraphics[width=7cm]{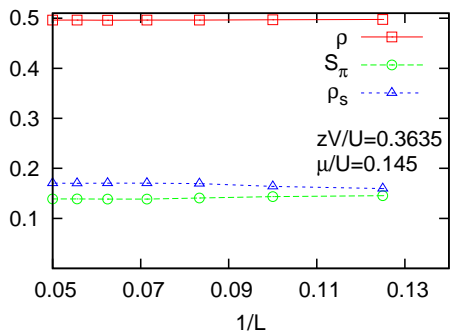}}
\caption[]{\label{fig:extrapolation}(Color online) The extrapolation of
  the order parameters in the supersolid phase with $\rho<0.5$. In the
  thermodynamic limit, the staggered structure factor $S_{\pi}=0.14$,
  the superfluid density $\rho_s=0.17$ and the particle density
  $\rho=0.496$, which indicates a supersolid state less than the half
  filling.}
\end{figure}

As shown in the previous study \cite{sengupta2005}, doping \emph{holes}
into the CDW I state may not lead to supersolid state with $\rho<0.5$ on
a two-dimensional square lattice, contrary to the case of doping
\emph{particles}. The reason is that the CDW I state has two-fold
degeneracy. As long as enough particles are removed, the insulating
state becomes unstable towards the formation of domain walls between the
two degenerate states, but can still be stable against losing \emph{one}
particle. This explains the particle density jump across the first-order
phase boundary between the superfluid and CDW I phases and that between
the supersolid and CDW II phases.  However this argument is invalid for
$V$ close to $t$ that is around the tip of the CDW I lobe, where we show
below that the kinetic energy gain by doping \emph{one} hole into the
CDW I state is also linear in $t$, that can cause instability of CDW I
state towards a supersolid state without the formation of domain
walls. Suppose that one particle is taken from a CDW I state, the hole
leaving behind moves in an effective staggered potential, roughly
speaking, $0$ in one sublattice and $(z-2)V$ in the other.  Solving this
single particle problem, one estimates the kinetic energy gain $\Delta
K$ is
\begin{eqnarray}
\label{eq:2}
\Delta K=-tz\left[\frac{V}{2t} ( 1- \frac{2}{z}) +
  \sqrt{1+ \left( \frac{V}{2t} \right)^2( 1- \frac{2}{z}
    )^2}\right]^{-1},
\end{eqnarray}
which is about $-2.88t$ in the cubic lattice near the tip of CDW I lobe
where $V\sim 2.4t$ (see Fig.~\ref{fig:phasediagram}). As a consequence,
a supersolid phase with $\rho<0.5$ occurs, whose boundary is plotted in
the inset of Fig.~\ref{fig:phasediagram}. To confirm, we also
extrapolate the static structural factor, superfluid order and particle
density at point $zV=0.3635U$ and $\mu=0.145$ to the thermodynamic limit
(see Fig.~\ref{fig:extrapolation}), which indicates a supersolid state
with the particle density less than half filling. For more information,
one can refer to next section where we shall examine the order
parameters in details.

In the presence of a kinetic term, the vertical boundary at $zV\sim U$
between Mott and CDW II states does not yet split but moves slightly to
the CDW side. On this boundary, the free energies of both states are
equal. However, they have different symmetries and are separated from
each other thermodynamically. The transition between these two
insulating states is of the first-order, which is similar to the
conventional liquid-solid phase transition. Further doping particles
into the Mott and CDW II states leads to the reentrance of superfluid
and supersolid states, respectively, with density $\rho>1$.

Because the particles or holes can hop between nearest neighbors on a
Mott background, which leads to a kinetic energy gain linear in $t$,
however they can only hop among next nearest neighbors in a staggered
CDW background that gives rise to a hopping energy proportional to
$t^2/V$. Then, as $t$ increases the Mott region shrinks much faster than
the CDW II region, which results in the mismatch of phase boundary
between them as $zV\sim U$(see Fig.~\ref{fig:phasediagram}). Since
doping the Mott state with holes leads to a superfluid state, one
expects a boundary between the superfluid state and the CDW II state in
the critical region at the lower end of the vertical boundary. This
phase boundary is again of the first-order. At the upper end, the
extension of the vertical boundary separates the superfluid and
supersolid phases in the way of the second-order phase transition.

\section{Order parameters}
\label{sec:order-parameters}
\begin{figure}[htpb]
\includegraphics[width=7cm]{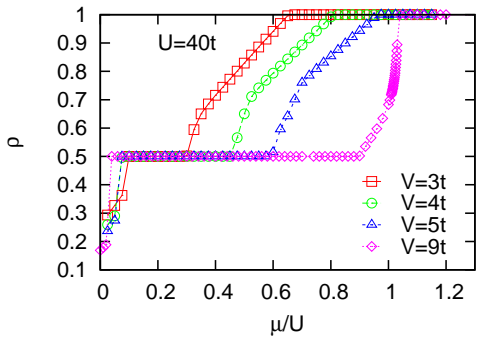}
\caption{(Color online) The average particle density $\rho$ as a
  function of chemical potential $\mu$ for different $V$ and
  $U=40t$. The first plateau at $\rho=0.5$ corresponds to the CDW I
  phase, and the second one corresponds to the Mott phase for
  $V/t=3,4,5$ and the CDW II phase for $V/t=9$. The system size is
  $12\times12\times12$.}
\label{fig:density}
\end{figure}
\begin{figure}[htpb]
\centering
\includegraphics[width=7.cm]{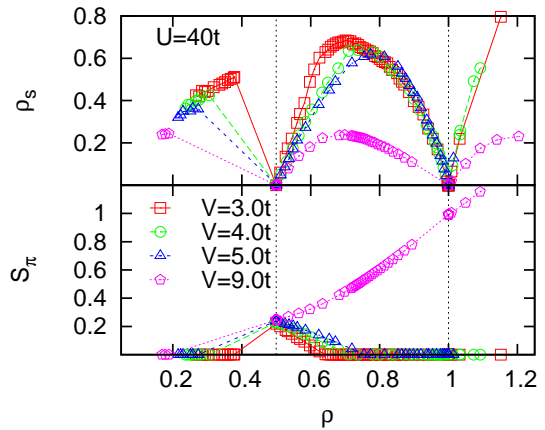}
\caption{(Color online) Superfluid density $\rho_s$ and staggered
  structure factor $S_{\vec{\pi}}$ as functions of the particle density
  $\rho$ at $U/t=40$. Note that the plateaus of constant $\rho$ in
  Fig.~\ref{fig:density} are only single points here.}
\label{fig:sf}
\end{figure}

In this section we give the details of the parameters $\rho$, $\rho_s$
and $S_{\vec{\pi}}$ for different $\mu$ and $V$.  Fig.~\ref{fig:density}
is the density profile as we vary the chemical potential for several
fixed values of $V$, where the plateaus correspond to the incompressible
states, i.e., the Mott, CDW I and II states, with vanishing isothermal
compressibility $\kappa_T\equiv \rho^{-2}\partial \rho/\partial
\mu$. Note that these insulating states correspond to two single points
$\rho=0.5$ and $\rho=1$ in Fig.~\ref{fig:sf}, which is the plot of
superfluid density $\rho_s$ and $S_{\vec{\pi}}$ as functions of particle
density $\rho$.

For $\rho<0.5$, the particles Bose condense to form a pure superfluid
state with a nonzero $\rho_s$ but vanishing $S_{\vec{\pi}}$ as shown in
Fig.~\ref{fig:sf}. When the particle density reaches a commensurate
value $\rho=0.5$, a plateau appears implying $\kappa_T=0$ which
corresponds to the incompressible CDW I state with translational
invariance broken. This transition is of the first-order since the
particle density, as a first-order derivative of free energy with
respect to the chemical potential, is discontinuous. This is also
reflected in Fig.~\ref{fig:sf} as the fact that a segment of particle
density below the half filling is inaccessible, and $S_{\vec{\pi}}$
jumps at $\rho=0.5$ to a finite value from zero.  As the particle
density exceeds $0.5$, $\kappa_T$ becomes a finite positive value
again. At the same time the superfluidity appears continuously upon the
CDW I state to form a supersolid. The corresponding transition is of the
second-order, since the particle density $\rho$ is continuous but
$\kappa_T$ jumps from zero to a finite value.

Between the two plateaus at $\rho=0.5$ and $\rho=1$, the slope of
$\rho(\mu)$ curve for $V/t=3,4$ and $5$, i.e. $zV<U$, shows another
jump, implying that $\kappa_T$ is discontinuous and a second-order phase
transition from the supersolid to superfluid phase occurs.  This
transition is manifested in Fig.~\ref{fig:sf}, where $\rho_s$ keeps
finite in the whose region $0.5<\rho<1.0$, but $S_{\vec{\pi}}$ vanishes
at some critical values in between. Further increasing the chemical
potential, the system enters into the Mott phase, which corresponds to
the plateau with $\rho=1$ in Fig.~\ref{fig:density}, through a
second-order phase transition. At the transition point $\rho_s$ vanishes
as shown in Fig.~\ref{fig:sf}.

For $V=9t$, i.e., $zV/U>1$, the second plateaus in
Fig.~\ref{fig:density} corresponds to the CDW II state, into which the
system enters directly from the supersolid phase as $\mu$ is
increased. The transition is a first-order one as shown in
Fig.~\ref{fig:density} as the jump of particle density. It is also
reflected in Fig.~\ref{fig:sf} where a segment of particle density
$\rho$ is not accessible before $\rho$ reaches 1.

\begin{figure}[htbp]
\centerline{\includegraphics[width=8.0cm]{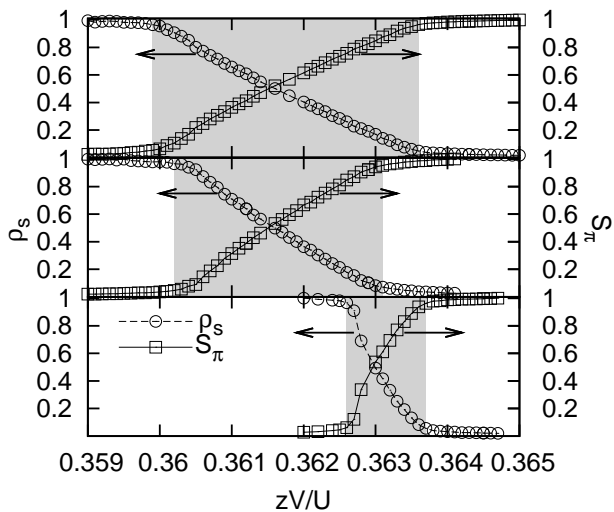}}
\caption[]{\label{fig:fixedmu} The normalized order parameters
  $\tilde{\rho}_s$ and $\tilde{S}_{\vec{\pi}}$ as functions of
  $zV/U$. The upper, middle, and bottom
  panels correspond to $\mu/U=0.160$, $0.155$ and $0.145$,
  respectively. The supersolid phases occur in shaded areas. The lattice
  size is $12\times12\times12$.}
\end{figure}

\begin{figure}[htbp]
\centerline{\includegraphics[width=8cm]{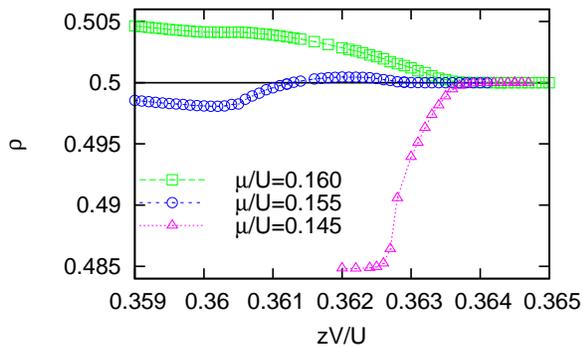}}
\caption[]{\label{fig:fixmurho} The particle density vs. $zV/U$ for
  $\mu/U=0.160$, $0.155$ and $0.145$.}
\end{figure}

In Fig.~\ref{fig:fixedmu}, we plot the normalized order parameters
$\tilde{\rho}_s\equiv\rho_s/\rho_{s,max}$ and
$\tilde{S}_{\vec{\pi}}\equiv S_{\vec{\pi}}/S_{\vec{\pi},max}$ as
functions of $zV/U$ around the CDW I lobe, where $\rho_{s,max}\approx
0.6$ and $S_{\vec{\pi},max}\approx 0.2$. In the plot, we take three
characteristic values of the chemical potential $\mu/U=0.160$, $0.155$
and $0.145$ from the top to the bottom panel.  In the shaded areas, when
$V$ increases, the superfluid density $\rho_s$ decreases while the
static structure factor $S_{\vec{\pi}}$ increases, and both $\rho_s$ and
$S_{\vec{\pi}}$ are nonzero in this region indicating a supersolid
state. The supersolid area becomes more and more narrow as $\mu$
decreases, until finally shrinks to a point in the $zV/U$ axis when
$\mu/U\sim 0.140$, which implies that two second-order phase boundaries
merge into one first-order phase boundary across which the order
parameters $\rho_s$ and $S_{\vec{\pi}}$ are both discontinuous. The
corresponding density profiles are plotted in Fig.~\ref{fig:fixmurho},
where we observe that the particle density is always smaller than $0.5$
for $\mu/U=0.145$ corresponding to the bottom panel in
Fig.~\ref{fig:fixedmu} until it enters into the CDW I phase, that
implies the supersolid state can exist below the half filling. For
$\mu/U=0.160$, the situation is different where $\rho$ decreases as $V$
is increasing, which can be intuitively attributed to the loss of
effective chemical potential due to nearest neighbor interaction $V$ in
the mean-field level. The case of $\mu/U=0.155$ shows an intermediate
behavior, for which the particle number first decreases, then increases
slightly larger than $0.5$ and finally reaches the CDW I state.

\section{Conclusion}
\label{sec:conclusion}

In this article, we present the global phase diagram
(Fig.~\ref{fig:phasediagram}) of the 3D extended Bose Hubbard model.
The EBH model exhibits four kinds of ground states, including (1) the
Mott state without breaking any symmetry, (2) the CDW I and II states
with translational symmetry broken, (3) the superfluid with U(1) gauge
symmetry broken, and (4) the supersolid with both symmetries broken. By
using the QMC simulation as well as other analytical tools, we identify
the transition type between these phases. Among them, the first-order
phase boundary includes those between the superfluid and CDW (I and II)
states, and between the Mott and CDW II states. The other boundaries are
all continuous. The critical regions for $zV/U\sim 1$ and the tip of the
CDW I lobe are examined in detail. We demonstrate that in the present 3D
EBH model, the supersolid phase with $\rho<0.5$ can appear in a small
region near the CDW I lobe where the hopping amplitude $t$ is comparable
to the nearest neighbor interaction $V$. In this region, the general
``domain wall'' argument for the nonexistence of supersolid state with
$\rho<0.5$ is no longer applicable, since it is based on the assumption
of $t\ll V$.

\section{ACKNOWLEDGMENT}

We acknowledge the helpful discussions with M. Ma. This work is
supported by NSFC Grant No. 10904081, 10934008, 90922033, HKSAR RGC
Grant No. HKU 701009 and Chinese Academy of Sciences.

%

\end{document}